\documentstyle[epsfig]{aipproc}

\begin{document}
\title{GRB Repetition Limits 
\\ from Current BATSE Observations} 
\author{Jon Hakkila$^*$, Charles A. Meegan$^{\ddagger}$, Geoffrey N. 
Pendleton$^{\dagger}$, Michael S. Briggs$^{\dagger}$, John M.
Horack$^{\ddagger}$, Dieter H. Hartmann$^{\circ}$,
and Valerie Connaughton$^{\star}$}
\address{$^*$Mankato State University, Mankato MN  56002-8400 \\
$^{\ddagger}$NASA/Marshall Space Flight Center, Huntsville, AL 35824 \\
$^{\dagger}$University of Alabama in Huntsville, Huntsville, AL 35899 \\
$^{\circ}$Clemson University, Clemson, SC 29634 \\
$^{\star}$National Research Council at NASA/MSFC, Huntsville, AL 35824}

\maketitle

\begin{abstract}
Revised upper limits on gamma-ray burst repetition rates are found using the 
BATSE 3B and 4B catalogs. A statistical  repetition model is assumed in which 
sources burst at a {\it mean} rate but in which BATSE observes bursts randomly 
from each source.
\end{abstract}

\section*{Introduction}

Regardless of the gamma-ray burst distance scale, the question of whether
bursters repeat (e.g. \cite{Quashnock93} \cite{Wang95}) is of
great importance to our understanding of them: their energy 
production mechanisms are constrained by repetition. 
There is strong statistical evidence to indicate that bursters are not
observed to repeat on timescales of days to months (\cite{Tegmark96}
\cite{Hakkila96} \cite{Brainerd96}), but the recent identification of
four gamma-ray bursts close together in both location and time 
\cite{Connaughton97} has left this possibility open for at least some bursters.
Of equal interest is the possibility that they might repeat on timescales 
of years or longer, and that repetition might take some form where they
do not burst steadily.

Detection of repetition is complicated by burst localization errors
(do two bursts belong to the same source?)
and by incomplete angular sky exposure and limited trigger efficiency (what
fraction of source repetitions is observed?). The repetition question
can be studied using two observational parameters
\cite{Meegan95} \cite{Hakkila96}; $f$ and $\langle\nu\rangle$. The variable $f$ 
represents the fraction of sources that produce more than one detected burst, 
while $\langle\nu\rangle$ represents the mean number of detected
bursts per source producing more than one detected burst.

\section*{Analysis}

{\bf Localization Errors.}
Localization errors are modeled as a combination in quadrature of a 
$1.6^{\circ}$ systematic error and a statistical error. The statistical error 
is a  combination of a fluence-dependent mean and a fluence-dependent random 
component obtained from the BATSE 4B Catalog.

{\bf Sky Exposure.}
The sky exposure used is that listed in the BATSE 4B Catalog. It should be 
noted that this exposure is only valid for non-overwriting bursts, so that 
overwriting bursts must be removed from the dataset.

{\bf Trigger Efficiency.}
The effects of trigger efficiency on repetition are not included in this
analysis. This is due in part to calibration work still in progress (see
Pendleton, Hakkila, \& Meegan, this conference) and in part to the
model-dependent way in which the repetition luminosity function enters into the
analysis (e.g. \cite{Band94} \cite{Hakkila96}).

{\bf The Burst Catalogs.}
For this analysis we choose to independently analyze the BATSE 3B 
\cite{Meegan96} and 4B catalogs. 
The catalogs have had all overwriting bursts removed, leaving 1060 3B bursts 
and 1554 4B bursts. We recognize that 4B bursts have 
been detected using several triggering energy channel combinations; this 
complicates the analysis in complex ways not modeled here. 
The 3B analysis does not suffer from this effect.

{\bf Clustering Statistic.}
The value of the Two-Point Angular Correlation Function found within
$7.2^{\circ}$ [TPACF($\theta \leq 7.2^{\circ}$)] is 
used as the clustering test for this analysis. This angle has been found to 
contain an optimum signal for repeating bursts \cite{Vo94} due to the
overall distribution of burst localization probabilities.

{\bf Repetition Models.}
Two models are used; a Euclidean model and a model based on an Einstein-de
Sitter cosmology with the faintest bursts at redshift $z=2$ (Hereafter referred
to as the ES2 model). Both spatial models have only one free parameter: it is
assumed that bursters burst with a mean intrinsic repetition rate $R_{\rm int}$
(repetitions source$^{-1}$ year$^{-1}$), but that equal probabilities exist 
that any burster might be the next one to burst. By varying $R_{\rm int}$, 
various fractions of the bursts $f$ are found to belong to clusters composed of 
various numbers of detected bursts $\nu$ (having an average value 
$\langle\nu\rangle$), which corresponds to a mean observed repetition rate 
$R_{\rm obs}$. In the limit 
of very low mean intrinsic repetition rates ($R_{\rm int} \approx 0$), these 
models are essentially the same as non-repeating models, since no clusters of 
two or more bursts will be produced during the BATSE operating time. In 
Euclidean space, $R_{\rm int}$ is always less than $R_{\rm obs}$, since the 
overall exposure is less than unity.  In the case of the ES2 model, 
$R_{\rm int}$ is 
assumed to be constant in the comoving frame.  More distant ES2 bursters appear
to repeat at a slower rate due to time dilation, so that a higher $R_{\rm int}$ 
is needed to produce a Euclidean equivalent $R_{\rm obs}$.

{\bf Monte Carlo Simulations.}
Monte Carlo simulations are performed
while varying the free parameter $R_{\rm int}$. It 
is found from the Monte Carlo simulations that
\begin{equation}
R_{\rm obs} = 0.481 R_{\rm int;Euclidean} = 0.310 R_{\rm int;ES2}
\end{equation}

Thus, there is no real difference between the Euclidean and ES2 cosmological
models in terms of the repetition {\it observables}.

\begin{figure}[b!] 
\centerline{\psfig{file=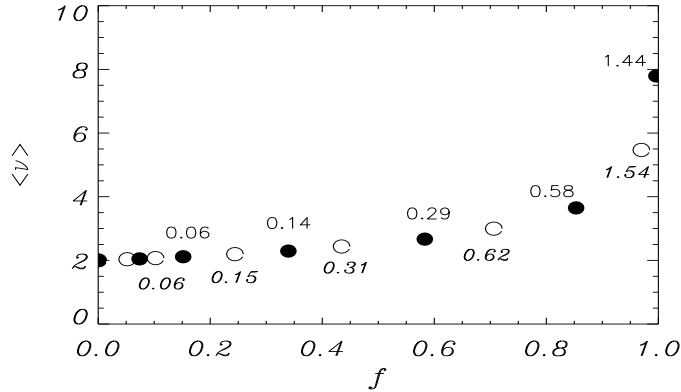,height=2.0in,width=3.5in}}
\vspace{12pt}
\caption{Correspondence between the mean observed repetition rate
$R_{\rm obs}$ and the Meegan $f$ and $\langle\nu\rangle$ parameters for the 
repetition modelin question. Specific rates $R_{\rm obs}$ (bursts source$^{-1}$ 
year$^{-1}$) are indicated for both the BATSE 3B (open circles; italic 
lettering) and 4B (closed circles; roman lettering) catalogs. The shape
of the $\langle\nu\rangle$ vs. $f$ relationships seen here is characteristic 
of this repetition model.}
\label{fig1}
\end{figure}

The Monte Carlo simulations allow $\langle\nu\rangle$ vs $f$ plots to be 
produced for both the BATSE 3B and 4B catalogs. These are presented in Figure 
\ref{fig1},along with the corresponding mean observed repetition rates. 
$R_{\rm obs}$ from $\langle\nu\rangle$ and $f$ differs between the two catalogs 
because the exposure differs. The 4B exposure is slightly better than the 3B 
exposure; if the two were identical the rates would coincide on the 
plots. If the rates were identical, then the longer 4B Catalog would produce 
more cluster composed of larger numbers of bursts. This effect is enhanced by 
the better 4B exposure, so that even more and larger clusters should be 
produced. It is therefore expected that repetition rates of the type described 
should be easier to identify in the 4B Catalog than in the 3B Catalog.

{\bf Repetition Limit Procedure.}
The BATSE 3B and 4B values of the TPACF within $7.2^{\circ}$ of one another 
[TPACF($\theta \leq 7.2^{\circ}$)] are slightly negative (the 3B value of 
$-0.070$ is $1.5\sigma$ from the isotropic value of zero, while the 
4B Catalog value of $-0.035$ is $1.1\sigma$
from isotropy), indicating a depletion of close burst pairs. 
Since models with nonzero repetition rates produce positive values of
this statistic, both BATSE catalogs are most consistent with no repetition.

We can also ask how consistent the results are with different mean observed
repetition rates $R_{\rm obs}$, so that we can get some idea concerning upper 
limits allowed on repetition rates. To do this, the 
TPACF($\theta \leq 7.2^{\circ}$) measurement (with associated error 
distribution; assumed Gaussian) at each $R_{\rm obs}$
can be compared to the minimum expected value of 
TPACF($\theta \leq 7.2^{\circ}$) $= 0$ at $R_{\rm obs} = 0$ (assuming that 
TPACF $= 0$ is the correct value). This sensitivity calculation
can provide some idea as to the maximum rate $R_{\rm obs}$ that would go 
undetected by the BATSE instrument.

\section*{Conclusions}

\begin{figure}[b!] 
\centerline{\epsfig{file=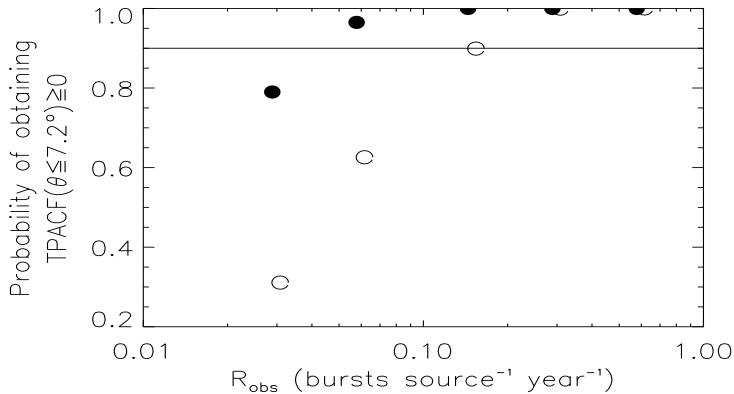,height=2.0in,width=3.5in}}
\vspace{12pt}
\caption{Probability of obtaining a value of TPACF($\theta \leq
7.2^{\circ}$) $\geq 0$ vs. mean observed repetition rate $R_{\rm obs}$. The plot
indicates that the larger 4B data set (closed circles) is more sensitive to 
detecting repetition of the type described, than the 3B data set (open 
circles). Unknown effects in the 4B, such as that of triggering on different 
energy channels, have not been taken into account in this analysis.}
\label{fig2}
\end{figure}

Figure \ref{fig2} indicates the probability that TPACF($\theta \leq 
7.2^{\circ}$) $\geq 0$ assuming different $R_{\rm obs}$ rates in the BATSE 3B 
and BATSE 4B  catalogs. At the 90\% confidence level, $R_{\rm obs} \leq 
0.15$ bursts source$^{-1}$ year$^{-1}$. The larger data set
of the BATSE 4B Catalog would place even greater constraints on the observed
repetition rate, if one considers the different triggering criteria used in
this catalog to be unimportant. For the 4B Catalog, it is unlikely that
$R_{\rm obs} \geq 0.04$ bursts source$^{-1}$ year$^{-1}$. 
This indicates that repetition is either absent or
that sources only repeat rarely on the timescales of the BATSE catalogs. These
90\% probabilities correspond to model-dependent limits on the mean intrinsic
repetition rates, which are indicated in Table \ref{table}.

\begin{table}
\caption{Limits on mean observed and intrinsic repetition rates (in units
of bursts source$^{-1}$ year$^{-1}$) for Euclidean and ES2 models. 
Limits indicate rates for which the probability is 90\% that 
TPACF($\theta \leq 7.2^\circ$) $\geq 0$, as opposed to the negative 
TPACF($\theta \leq 7.2^\circ$) values obtained from the BATSE 3B and 4B
catalogs.}
\label{table}
\begin{tabular}{||c|c|c|c||}
Catalog & $R_{\rm obs}$ Limit & $R_{\rm int;Euclidean}$ Limit & 
$R_{\rm int;RS2}$ Limit \\
\tableline
BATSE 3B & 0.15 & 0.31 & 0.48 \\
BATSE 4B & 0.05 & 0.10 & 0.16 \\
\end{tabular}
\end{table}

In a previous work, we \cite{Hakkila96} quoted similar types of limits on the
Euclidean mean intrinsic repetition rate, finding that $R_{\rm obs} \geq 0.05$ 
bursts source$^{-1}$ year$^{-1}$ was unlikely in the BATSE 3B Catalog. Our 
revised 3B 90\% upper limit is higher at $R_{\rm obs} = 0.31$ bursts 
source$^{-1}$ year$^{-1}$, in part due to a more
direct definition of our statistical significance, and in part indicating that
BATSE is not as sensitive to detecting repetition as a result of the improved 
sky exposure analysis. 
The new sky exposure requires that overwrites be excluded; with fewer
bursts, higher repetition rates are needed to be detected above the random
background. The new exposure is also higher than that estimated before,
indicating that more of any intrinsic repetitions should be detected. By
selecting fewer bursts from sources, cluster sizes tend to be smaller. Thus 
$\langle\nu\rangle$ is smaller for a given repetition rate, and is more 
difficult to detect.

Our analysis technique could still be more sensitive to detecting repetition 
than a recent technique that combines
localization probabilities from multiple experiments to obtain smaller
localization regions \cite{Kippen98}. That technique limits its 
database to bursts
detected by both experiments, so that the sky exposure and trigger efficiency
for the combined experiment must then be found by multiplying these {\it
detection} probabilities together. We have shown here that limits on
repetition {\it rates} depend both on localization uncertainty and sky 
exposure. 

Our repetition model allows that some observed clusters can be large, and in
fact suggests that is it possible for the October 1996 burst ``cluster'' 
\cite{Connaughton97} to have
come from one source. From the upper limit on the repetition rate described
above, it is expected that at most 29 clusters of 3 bursts and 4 clusters of 4
bursts should be observed in the BATSE 3B Catalog. The tighter limits on 4B
Catalog repetition, however, indicate that only as many as 19 clusters of 3
bursts and 1 cluster of 4 bursts are expected.


\begin{references}
\bibitem{Band94}Band, D. J., 1994, {\it ApJ},  {\bf 422}, L75.
\bibitem{Brainerd96}Brainerd, J. J., 1996, {\it ApJ}, {\bf 473}, 974.
\bibitem{Connaughton97}Connaughton, V. et al., 1997, in {\it Proc. 18th Texas 
Symposium}, ed. A. Olinto, in press.
\bibitem{Hakkila96}Hakkila, J., et al., 1996, in {\it Gamma-Ray Bursts}, 
ed. C. Kouveliotou, M. S. Briggs, \& G. J. Fishman, (AIP: New York),  p. 392.
\bibitem{Kippen98}Kippen, R. M. et al., 1998, (this conference).
\bibitem{Meegan95}Meegan, C. A. et al., 1995, {\it ApJ}, {\bf 446}, L15.
\bibitem{Meegan96}Meegan, C. A. et al., 1996, {\it ApJS}, {\bf 106}, 65.
\bibitem{Quashnock93}Quashnock, J. M., \& Lamb, D. Q., 1993, {\it MNRAS},
{\bf 265}, L45.
\bibitem{Tegmark96}Tegmark, M. et al., 1996, {\it ApJ}, {\bf 466}, 757.
\bibitem{Vo94}Vo, V., 1994, Master's Thesis, Mankato State University, Mankato,
MN.
\bibitem{Wang95}Wang, V. \& Lingenfelter, R. E., 1995, {\it ApJ}, {\bf 441}, 
747.
\end{references}
\end{document}